\documentclass[a4paper,11pt]{article}
\usepackage{pos}

\usepackage{braket}
\usepackage{bm,bbm}

\newcommand\beq{ \begin{eqnarray} }
\newcommand\eeq{ \end{eqnarray} }

\title{Computing theta-dependent mass spectrum of the 2-flavor Schwinger model in the Hamiltonian formalism}
\ShortTitle{Theta-dependent mass spectrum of the 2-flavor Schwinger model}

\author*[a,b]{Akira Matsumoto}
\author[a,b]{Etsuko Itou,}
\author[a]{Yuya Tanizaki}

\affiliation[a]{
Yukawa Institute for Theoretical Physics, Kyoto University,\\ 
Kitashirakawa Oiwakecho, Sakyo-ku, Kyoto 606-8502 Japan}

\affiliation[b]{
Interdisciplinary Theoretical and Mathematical Sciences Program (iTHEMS), RIKEN,\\ 
2-1 Hirosawa, Wako, Saitama 351-0198 Japan}

\emailAdd{itou(at)yukawa.kyoto-u.ac.jp}
\emailAdd{akira.matsumoto(at)yukawa.kyoto-u.ac.jp}
\emailAdd{yuya.tanizaki(at)yukawa.kyoto-u.ac.jp}

\abstract{
We compute the $\theta$-dependent mass spectrum of the 2-flavor Schwingr model 
using the tensor network (DMRG) in the Hamiltonian formalism.
The pion and the sigma meson are identified as stable particles of the model 
for nonzero $\theta$ whereas the eta meson becomes unstable. 
The meson masses are obtained from the one-point functions, 
using the meson operators defined by diagonalizing the correlation matrix 
to deal with the operator mixing. 
We also compute the dispersion relation directly 
by measuring the energy and momentum of the excited states, 
where the mesons are distinguished by the isospin quantum number. 
We confirmed that the meson masses computed by these methods agree with each other 
and are consistent with the calculation by the bosonized model. 
Our methods are free from the sign problem 
and show a significant improvement in accuracy compared to the conventional Monte Carlo methods.
Furthermore, at the critical point $\theta = \pi$, the mesons become almost massless, 
and the one-point functions reproduce the expected CFT-like behavior.
This talk is based on the paper~\cite{Itou:2024psm}.}

\FullConference{The 41st International Symposium on Lattice Field Theory (LATTICE2024)\\
 28 July - 3 August 2024\\
Liverpool, UK\\}

\begin{document}

\begin{flushright}
    YITP-25-08, RIKEN-iTHEMS-Report-25
\end{flushright}

\maketitle

\section{Introduction and summary}

The lattice Monte Carlo simulation has played an essential role in uncovering various nonperturbative phenomena of QCD.
One of the important applications is predicting the hadron mass spectrum, 
where experimental results can be reproduced only from a few input parameters~\cite{FlavourLatticeAveragingGroupFLAG:2021npn}.
On the other hand, the sign problem has been a long-standing obstacle 
to studying systems with the chemical potential, the topological term, etc.
To circumvent this problem, numerical methods in the Hamiltonian formalism, 
such as quantum computing and tensor networks, are recently developed as complementary approaches.
These new methods directly handle approximated wave functions and do not rely on important sampling, 
which will enable us to access new information that is hard to obtain by the conventional Monte Carlo method.

In ref.~\cite{Itou:2023img}, we developed three distinct methods to compute the mass spectrum 
in the Hamiltonian formalism and demonstrated them by studying the 2-flavor Schwinger model at $\theta = 0$.
There are three composite particles characterized by quantum numbers of the isospin $J$, parity $P$, and $G$-parity $G$, 
and they are called the pion $J^{PG}=1^{-+}$, sigma meson $0^{++}$, and eta meson $0^{--}$.
We obtained the promising results of their mass spectrum at $\theta = 0$, 
which motivates us to extend the work to $\theta \neq 0$.
However, the nonzero $\theta$ angle explicitly breaks $P$ and $G$, 
so that the eta meson cannot be distinguished from the other singlet states by these symmetries.
Indeed, the eta meson becomes unstable for $\theta \neq 0$ due to the $\sigma-\eta$ mixing and $\eta \to \pi\pi$ decay.
Furthermore, the analytic study on the bosonized model~\cite{Coleman:1976uz} predicts 
that the gap of the system, namely the pion mass, decreases as $\theta$ approaches $\pi$, 
which will make our numerical method more costly.
Thus, we need to improve the methods and deal with these subtleties 
so that we can investigate the $\theta$-dependent mass spectrum by the first-principles calculation.
In ref.~\cite{Itou:2024psm}, we use two independent schemes; the improved one-point-function and the dispersion-relation schemes.

\begin{figure}[h]
    \centering
    \includegraphics[scale=0.65]{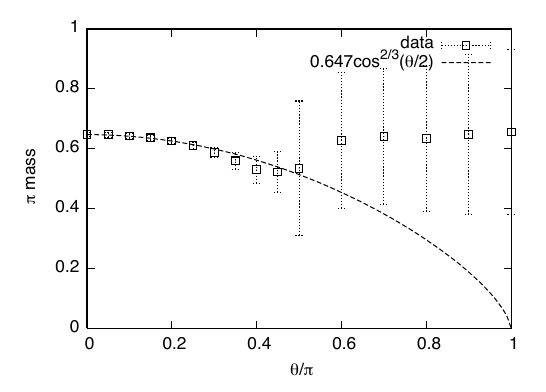} 
    \includegraphics[scale=0.4]{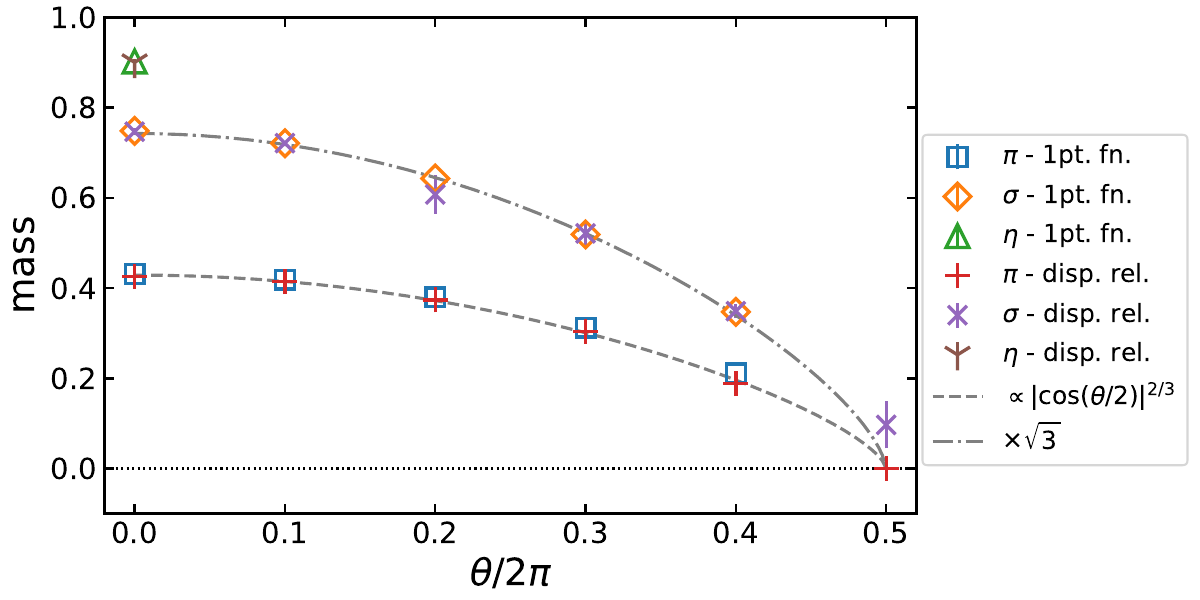} 
    \caption{\label{fig_mass_comp}
    (Left) The Monte Carlo result of the pion mass is taken from ref.~\cite{Fukaya:2003ph}.
    The reweighting method was used to include the effect of the $\theta$ angle.
    (Right) The meson masses obtained by the improved one-point-function and dispersion-relation schemes 
    are plotted against $\theta/2\pi$.
    The dashed and dash-dotted curves denote the analytic results by the bosonization, 
    $M_{\pi}(\theta)=M_{\pi}(0)|\cos(\theta/2)|^{2/3}$ and $M_{\sigma}(\theta)=\sqrt{3}M_{\pi}(\theta)$, respectively.
    Here the overall coefficient $M_{\pi}(0)$ is determined by averaging the results of the two schemes at $\theta = 0$.}
\end{figure}

In figure~\ref{fig_mass_comp}, we compare our numerical results 
with the Monte Carlo calculation of the $\theta$-dependent pion mass by Fukaya and Onogi in 2003~\cite{Fukaya:2003ph}.
Even with careful treatment of the reweighting factor, 
the Monte Carlo result suffers from the severe sign problem in the large $\theta$ region (left panel).
On the other hand, our results (right panel) of both the two schemes are precise even for large $\theta$ and consistent with each other, 
where the mass of not only the pion but also the sigma meson can be computed.
In particular, the dispersion-relation scheme has an advantage in studying the system heuristically 
when low-lying states are unknown.
The reason is that this scheme does not rely on any local operators, 
unlike the Monte Carlo method which requires appropriate ones depending on the target hadron.

In figure~\ref{fig_mass_comp}, we also compare the numerical results with the analytic calculation by the bosonization, 
where the pion mass is predicted as $M_{\pi}(\theta) \propto |m \cos (\theta/2)|^{2/3}$ 
when the fermion mass $m$ is small~\cite{Coleman:1976uz}.
Furthermore, the WKB-type approximation gives a specific mass ratio, 
$M_{\sigma}(\theta)=\sqrt{3}M_{\pi}(\theta)$, between the pion and sigma meson.
These analytic results are depicted in figure~\ref{fig_mass_comp}, 
normalized by the numerical data of the pion mass at $\theta = 0$.
The numerical results agree with the analytic prediction, 
which indicates that the bosonization gives almost the correct answer for the wide range of $\theta$.

This proceeding paper is organized as follows.
In section~\ref{sec_method}, we review the lattice Hamiltonian formalism and the numerical method.
In section~\ref{sec_1pt_func}, we explain the improved one-point-function scheme and show the simulation results.
In section~\ref{sec_disp_rel}, we consider the dispersion-relation scheme and show the result of the spectrum.
Section~\ref{sec_discussion} is devoted to the discussion.

\section{Lattice Hamiltonian and calculation strategy}
\label{sec_method}

Let us briefly review the lattice Hamiltonian formalism for the numerical computation. 
We consider the 2-flavor Schwinger model on an open interval.
In this case, the Gauss' law condition can be solved explicitly, 
so that the Hamiltonian is described only by fermions after the gauge fixing.
We adopt the staggered fermion~\cite{Kogut:1974ag, Susskind:1976jm} for the lattice discretization 
and apply the Jordan-Wigner transformation to obtain the spin Hamiltonian with a finite-dimensional Hilbert space, 
\begin{align}
    H= & \frac{g^2a}{8}\sum_{n=0}^{N-2}
    \left[\sum_{f=1}^{N_{f}}\sum_{k=0}^{n}\sigma_{f,k}^{z}
    +N_{f}\frac{(-1)^{n}+1}{2}+\frac{\theta}{\pi}\right]^2 \nonumber \\
     & -\frac{i}{2a}\sum_{n=0}^{N-2}\left(
    \sigma_{1,n}^{+}\sigma_{2,n}^{z}\sigma_{1,n+1}^{-}
    +\sigma_{2,n}^{+}\sigma_{1,n+1}^{z}\sigma_{2,n+1}^{-}
    -\mathrm{h.c.}\right)
    +\frac{m_{\mathrm{lat}}}{2}\sum_{f=1}^{N_f}\sum_{n=0}^{N-1}(-1)^{n}\sigma_{f,n}^z,
    \label{eq_H_spin}
\end{align}
which is convenient for applying quantum computing or tensor network algorithms.
Here the lattice fermion mass $m_{\mathrm{lat}}$ is related to the mass $m$ of the continuum theory 
by $m_{\mathrm{lat}}:=m-N_f g^2 a/8$ so that the $\mathbb{Z}_2$ discrete chiral symmetry could be recovered 
in the chiral limit~\cite{Dempsey:2022nys}.

We use the density-matrix renormalization group (DMRG)~\cite{White:1992zz, White:1993zza, Schollw_ck_2005, Schollw_ck_2011} 
to study the Hamiltonian~\eqref{eq_H_spin}.
The ground state is variationally obtained as a matrix product state (MPS) that minimizes the energy as a cost function, 
applying the low-rank approximation by the singular-value decomposition.
In the DMRG of the 2-flavor Schwinger model, potential difficulty arises from the entanglement property.
The MPS is an efficient representation of low-energy states of the $(1+1)$d gapped system 
because the entanglement entropy is independent of the system size and accommodated by the constant bond dimension.
However, the gap of our model decreases as $\theta$ increases 
and the system eventually becomes nearly conformal at $\theta = \pi$~\cite{Coleman:1976uz, Dempsey:2023gib}, 
where the entanglement entropy scales as $S_{\mathrm{EE}} \sim (c/3)\log N$ with the lattice size $N$ and the central charge $c=1$.
Thus, the required bond dimension is no longer constant but $D \sim N^{c/3}$.
In our calculation, we take at most $N=320$ and $a=0.25$, 
which results in the bond dimension $D \lesssim 1400$ at $\theta = \pi$.
This calculation is still doable using a PC cluster or a single-node supercomputer.

In the following analyses, we fix the gauge coupling to $g = 1$ 
to measure the energy scale in this unit of the mass dimension $1$ and set the fermion mass to $m = 0.1$.
The topological angle $\theta$ is varied from $0$ to $\pi$ in increments of $0.1/2\pi$.
We used the C++ library of ITensor~\cite{itensor} to perform the tensor network calculation.

\section{Improved one-point-function scheme}
\label{sec_1pt_func}

In this section, we discuss the basic strategy of the improved one-point-function scheme 
and present the numerical results for the 2-flavor Schwinger model.
This scheme utilizes the boundary as a source of excitation from the thermodynamic ground state.
The key point is that the one-point function decays exponentially as 
$\braket{\mathcal{O}(x)} \sim e^{-Mx}$ with the distance $x$ from the boundary, 
where $M$ is the mass of the lightest meson with the same quantum number as the operator $\mathcal{O}$.
We measure $\braket{\mathcal{O}(x)}$ and then compute the effective mass to obtain the meson mass $M$ by fitting.

For $\theta \neq 0$, the meson operators should be defined taking the operator mixing into account.
We deal with the operator mixing by using the correlation functions so that the operators can be defined systematically.
However, $\theta = \pi$ is a special point where the pion and sigma meson are almost massless.
In this case, the one-point functions are no longer the exponential type but CFT-like.
Thus, we compare them with the analytic calculation of the WZW CFT, instead of computing the meson masses.

In the following subsections, 
we first discuss the operator mixing and define the meson operators in section~\ref{subsec_op_mixing}.
Next, we show the main numerical results for $0 \leq \theta < \pi$ in section~\ref{subsec_1pt_main}.
Finally, the one-point functions at $\theta = \pi$ are investigated in section~\ref{subsec_1pt_CFT}.

\subsection{Operator mixing}
\label{subsec_op_mixing}

Here we explain the method to resolve the operator mixing by using the information of the correlation functions.
We then define the meson operators to measure the appropriate one-point functions for $\theta \neq 0$.
The meson operators are formally written by the fermion bilinear operators 
transformed by the $\theta$-depedent axial rotation as 
\begin{equation}
    \pi_{a} = -i\bar{\psi}\, e^{i\frac{\theta}{2}\gamma^5} \gamma^{5} \tau_{a} \psi,
    \qquad
    \sigma = \bar{\psi}\, e^{i\left(\frac{\theta}{2}+\omega(\theta)\right)\gamma^5} \psi,
    \qquad
    \eta = -i\bar{\psi}\, e^{i\left(\frac{\theta}{2}+\omega(\theta)\right)\gamma^5} \gamma^{5} \psi,
    \label{eq_axial_rot}
\end{equation}
where the extra rotation $\omega(\theta)$ comes from the effect of the $\sigma-\eta$ mixing.
Since the exact mixing angle is not known, we determine $\omega(\theta)$ numerically.

The mixing angle can be extracted from the correlation matrix of the scalar and pseudo-scalar operators, 
\begin{equation}
    \boldsymbol{C}_{\pm}(x,y)=
    \begin{pmatrix}
        \Braket{S_{\pm}(x)S_{\pm}(y)}_{c} & \Braket{S_{\pm}(x)PS_{\pm}(y)}_{c}\\
        \Braket{PS_{\pm}(x)S_{\pm}(y)}_{c} & \Braket{PS_{\pm}(x)PS_{\pm}(y)}_{c}
    \end{pmatrix}, 
\end{equation}
where $S_{+} \leftrightarrow \bar{\psi}\psi$ and $PS_{+} \leftrightarrow -i\bar{\psi}\gamma^5\psi$ 
denote the iso-singlet operators on the lattice, and $S_{-}$ and $PS_{-}$ are their iso-triplet counterparts.
The subscript $c$ means the connected part of the correlator.
Since $\boldsymbol{C}_{\pm}(x,y)$ is a real symmetric matrix, 
it can be diagonalized by an orthogonal matrix $R(\delta)$ with a rotation angle $\delta$.
Then $\delta$ corresponds to the mixing angle, and the eigenvalues are the correlation functions of the meson operators.
For example, the relation for the iso-singlet sector is given by 
\begin{equation}
    \boldsymbol{C}_{+}(x,y)=
    R(\delta_+)^{\mathrm{T}}
    \begin{pmatrix}
        \Braket{\sigma(x)\sigma(y)}_{c} & 0\\
        0 & \Braket{\eta(x)\eta(y)}_{c}
    \end{pmatrix}
    R(\delta_+).
\end{equation}
Using the matrix $R(\delta_{\pm})$, 
the appropriate meson operators $\pi(x)$, $\sigma(x)$, and $\eta(x)$ can be defined as 
\begin{equation}
    \begin{pmatrix}
        * \\ \pi(x)
    \end{pmatrix}
    =R(\delta_-)
    \begin{pmatrix}
        S_{-}(x) \\ PS_{-}(x)
    \end{pmatrix},
    \qquad 
    \begin{pmatrix}
        \sigma(x) \\ \eta(x)
    \end{pmatrix}
    =R(\delta_+)
    \begin{pmatrix}
        S_{+}(x) \\ PS_{+}(x)
    \end{pmatrix}.
    \label{eq_meson_op}
\end{equation}
Note that there is no mixing counterpart of the pion due to the absence of iso-triplet scalar particles.

\begin{figure}[tb]
    \centering
    \includegraphics[scale=0.35]{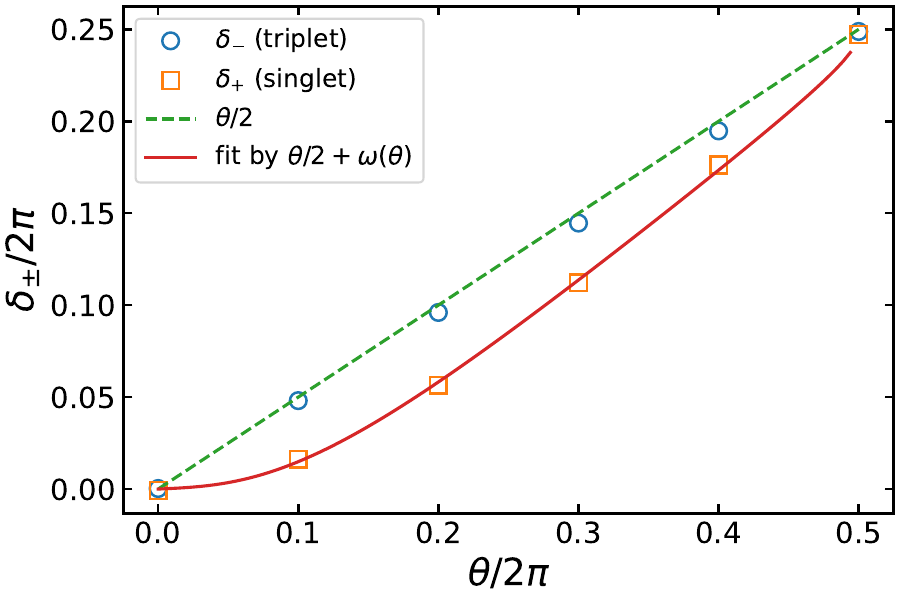}
    \caption{\label{fig_mixing} 
    The mixing angles $\delta_{-}$ of the triplet and $\delta_{+}$ of the singlets are plotted against $\theta/2\pi$.
    The dashed line depicts the expected linear behavior of $\theta/2$ for the triplet sector.
    The solid curve denotes the fitting result $\theta/2+\omega(\theta)$ for the singlet sector.}
\end{figure}

We compute the mixing angle $\delta_{\pm}$ on the lattice with the size $N=320$ and the spacing $a=0.25$, 
and the result is shown in figure~\ref{fig_mixing}.
It is confirmed that the mixing angle for the triplet sector is consistent with the trivial axial rotation 
$\delta_{-} = \theta/2$ whereas that for the singlet sector deviates from $\theta/2$ as expected.
The effect of the $\sigma-\eta$ mixing, $\omega(\theta)$, in eq.~\eqref{eq_axial_rot} can be analytically computed 
by the bosonization leaving a few unknown parameters\footnote{
See section~4.1.1 in ref.~\cite{Itou:2024psm} for a detailed discussion.}.
Indeed, the analytic form of $\theta/2 + \omega(\theta)$ fits the numerical result of $\delta_{+}$ well, 
as depicted by the solid curve in figure~\ref{fig_mixing}.
The agreement of the fitting result indicates consistency 
with the bosonized description of the $\sigma-\eta$ mixing at finite $m$.

\subsection{Meson mass obtained from the one-point function}
\label{subsec_1pt_main}

Next, we move on to the main result of the improved one-point-function scheme.
The meson mass is obtained from the exponential decay of the one-point function, 
which can be understood as an analogy to the wall-source method in Lattice QCD.
In Euclidean space, the spacial boundary is translationally invariant in the imaginary-time direction, 
which imposes zero-momentum projection as a wall source of mesons.
Thus, the asymptotic behavior of $\braket{\mathcal{O}(x)}$ should be $e^{-Mx}$ for the target meson of mass $M$.

To measure the nonzero one-point function, the boundary state, namely the source of the meson, 
must have the same quantum number as the target meson.
Thus, we control the boundary condition by attaching supplemental lattice sites, 
named ``the wings regime'', to both ends of the $1$d lattice, 
in which we assign a different fermion mass $m_{\mathrm{wings}}$ than the mass $m$ in the bulk.
For the iso-singlet mesons, we set the mass in the wings to a flavor-independent large constant, 
$m_{\mathrm{wings}} = m_0 \gg m$, resulting in the Dirichlet boundary.
As for the iso-triplet meson, we apply a flavor-asymmetric chiral rotation, 
$m_{\mathrm{wings}}=m_0 \, e^{\pm i\Delta\gamma^5}$, to implement the isospin-breaking effect.

\begin{figure}[tb]
    \centering
    \includegraphics[scale=0.35]{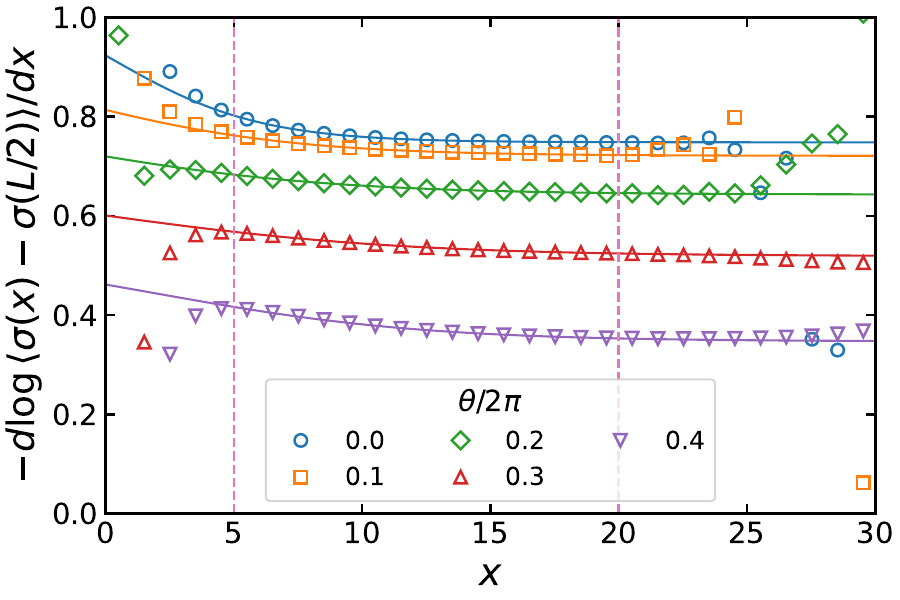}
    \quad
    \includegraphics[scale=0.35]{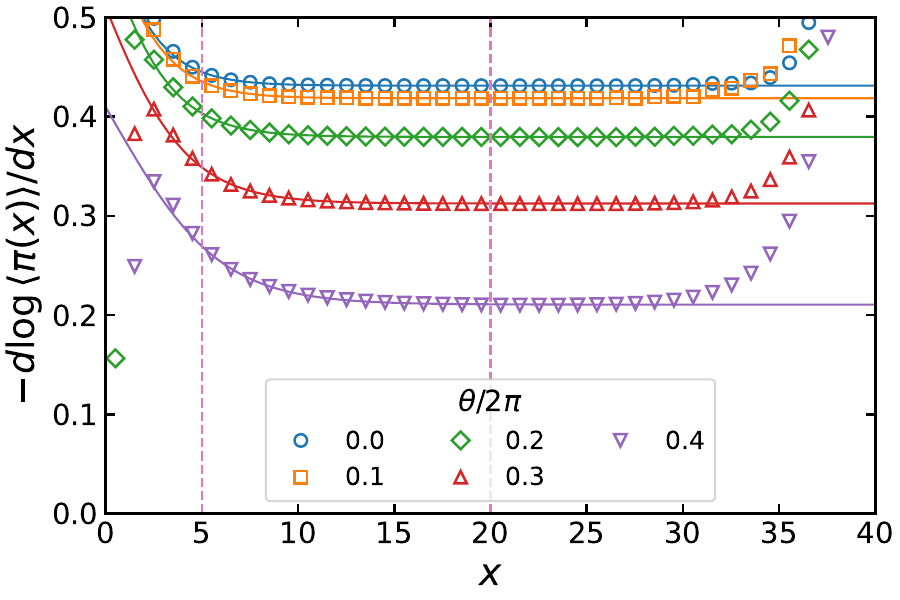}
    \caption{\label{fig_1pt_scheme} 
    The effective masses of the sigma meson (left) and pion (right) computed from their one-point functions 
    are plotted against the distance $x$ from the boundary.
    The solid curves depict the fitting results, where the fitting range is between the vertical dashed lines.}
\end{figure}

In this calculation, 
the lattice size of the bulk and the wings regime are $N=320$ and $N_{\mathrm{wings}} = 20$, 
respectively, with the spacing $a=0.25$.
We first generate the ground state and obtain the correlation matrix 
to define the meson operators as eq.~\eqref{eq_meson_op}.
Then we measure the one-point function $\braket{\mathcal{O}(x)}$ 
and compute the effective mass, $M_{\mathrm{eff}}(x) = -d\log \braket{\mathcal{O}(x)} / dx$, 
as shown in figure~\ref{fig_1pt_scheme}.
The effective mass is expected to be constant at long distances.
However, it is slightly curved by the contribution of the excited states, 
which cannot be ignored as the meson mass becomes small for large $\theta$.
Thus, we assume the one-point functions to be $\braket{\mathcal{O}(x)} \sim Ae^{-Mx} + Be^{-(M+\Delta M)x}$, 
incorporating the second-lowest state with the mass gap $M + \Delta M$.
We obtain the meson mass $M$ by fitting the effective mass, 
resulting in the fitting curves shown in figure~\ref{fig_1pt_scheme}.

\subsection{Nearly conformal behavior at \texorpdfstring{$\theta=\pi$}{theta=pi}}
\label{subsec_1pt_CFT}

\begin{figure}[h]
    \centering
    \includegraphics[scale=0.35]{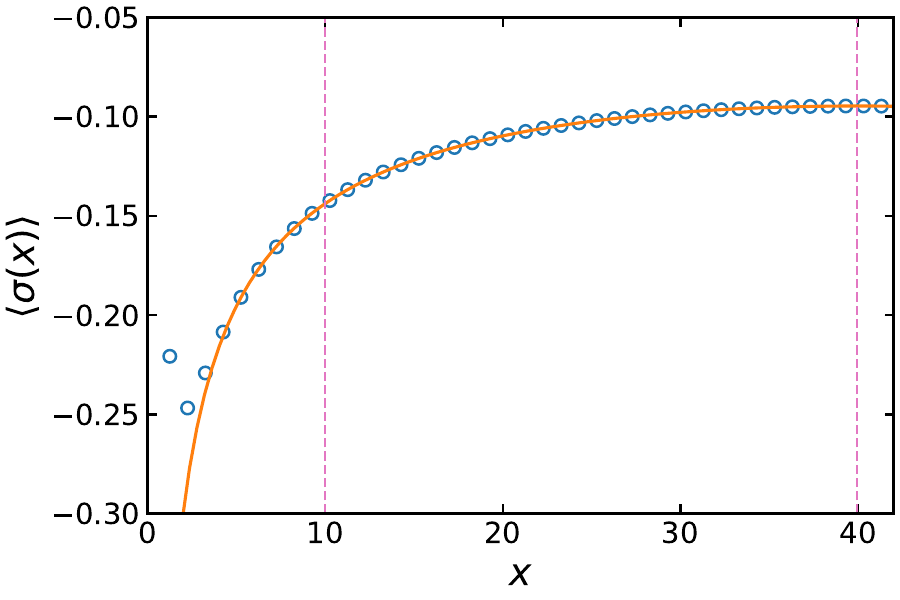}
    \quad
    \includegraphics[scale=0.35]{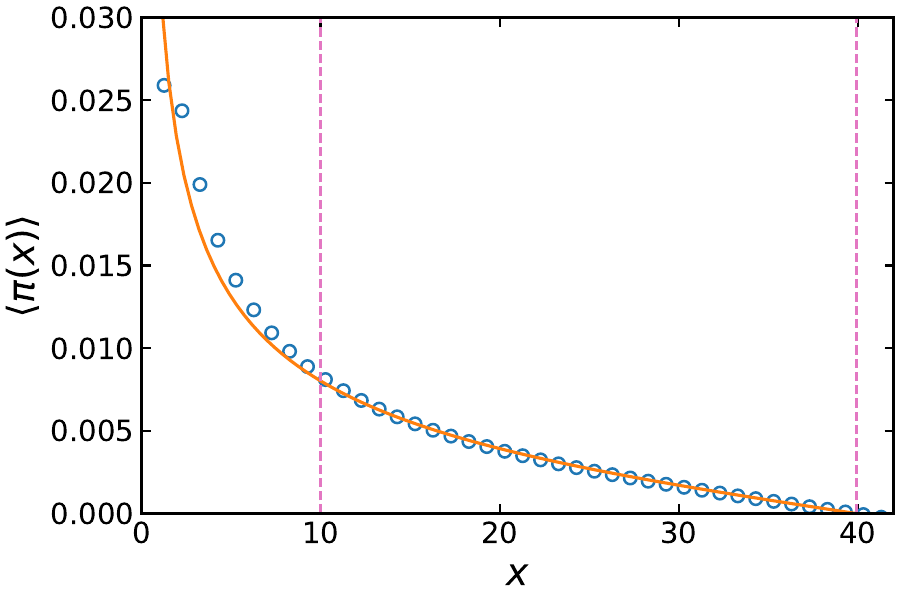}
    \caption{\label{fig_cft} 
    The one-point functions at $\theta = \pi$ of the sigma meson (left) and pion (right) are plotted against $x$.
    The fitting results based on the analytic forms~\eqref{eq_1pt_WZW} are depicted by the solid curves, 
    where the fitting range is between the vertical dashed lines.}
\end{figure}

At $\theta = \pi$, 
the 2-flavor Schwinger model approaches the nearly conformal theory~\cite{Coleman:1976uz, Dempsey:2023gib}.
In this case, the one-point functions are not the exponential type as shown in figure~\ref{fig_cft}.
Since the fitting ansatz in section~\ref{subsec_1pt_main} does not work there, 
the corresponding effective mass is absent in figure~\ref{fig_1pt_scheme}.

We compare the one-point functions at $\theta = \pi$ with the analytic calculation 
of the $\mathrm{SU}(2)_1$ WZW model on the finite interval $0 \leq x \leq L$, which predicts 
\begin{equation}
    \Braket{\sigma(x)} \propto -\frac{1}{\sqrt{\sin (\pi x/L)}},
    \qquad
    \Braket{\pi(x)} \propto \frac{\sin [\Delta(1-2x/L)]}{\sqrt{\sin (\pi x/L)}},
    \label{eq_1pt_WZW}
\end{equation}
for the sigma meson with the Dirichlet boundary and the pion with the isospin-breaking boundary\footnote{
See appendix~A in ref.~\cite{Itou:2024psm} for computational details.}.
The numerical data of the one-point functions can be fitted well by these analytic forms in the bulk region, 
as depicted by the solid curves in figure~\ref{fig_1pt_scheme}.
Thus, we numerically confirmed that the pion and sigma meson at $\theta = \pi$ are well described by the WZW CFT.
Even when the system becomes CFT-like, we can obtain consistent results of the one-point functions by the DMRG.

\section{Dispersion-relation scheme}
\label{sec_disp_rel}

Let us discuss the dispersion-relation scheme, a distinctive strategy in the Hamiltonian formalism.
The key idea of this scheme is to generate the energy eigenstates 
and identify the momentum excitations of each meson heuristically by the quantum numbers.
Then the dispersion relation $E = \sqrt{K^2+M^2}$ is obtained from their energy $E$ and momentum square $K^2$.
The low-energy excited states can be generated by the DMRG, modifying the Hamiltonian as 
\begin{equation}
    H_{\ell} = H + W\sum_{\ell^{\prime}=0}^{\ell-1}\ket{\Psi_{\ell^{\prime}}}\bra{\Psi_{\ell^{\prime}}},
    \label{eq_H_ell}
\end{equation}
where $\ell$ is the level of the target state 
and $W>0$ is a weight to impose the orthogonality~\cite{Wall_2012, Banuls:2013jaa}.


\begin{figure}[h]
    \centering
    \includegraphics[scale=0.3]{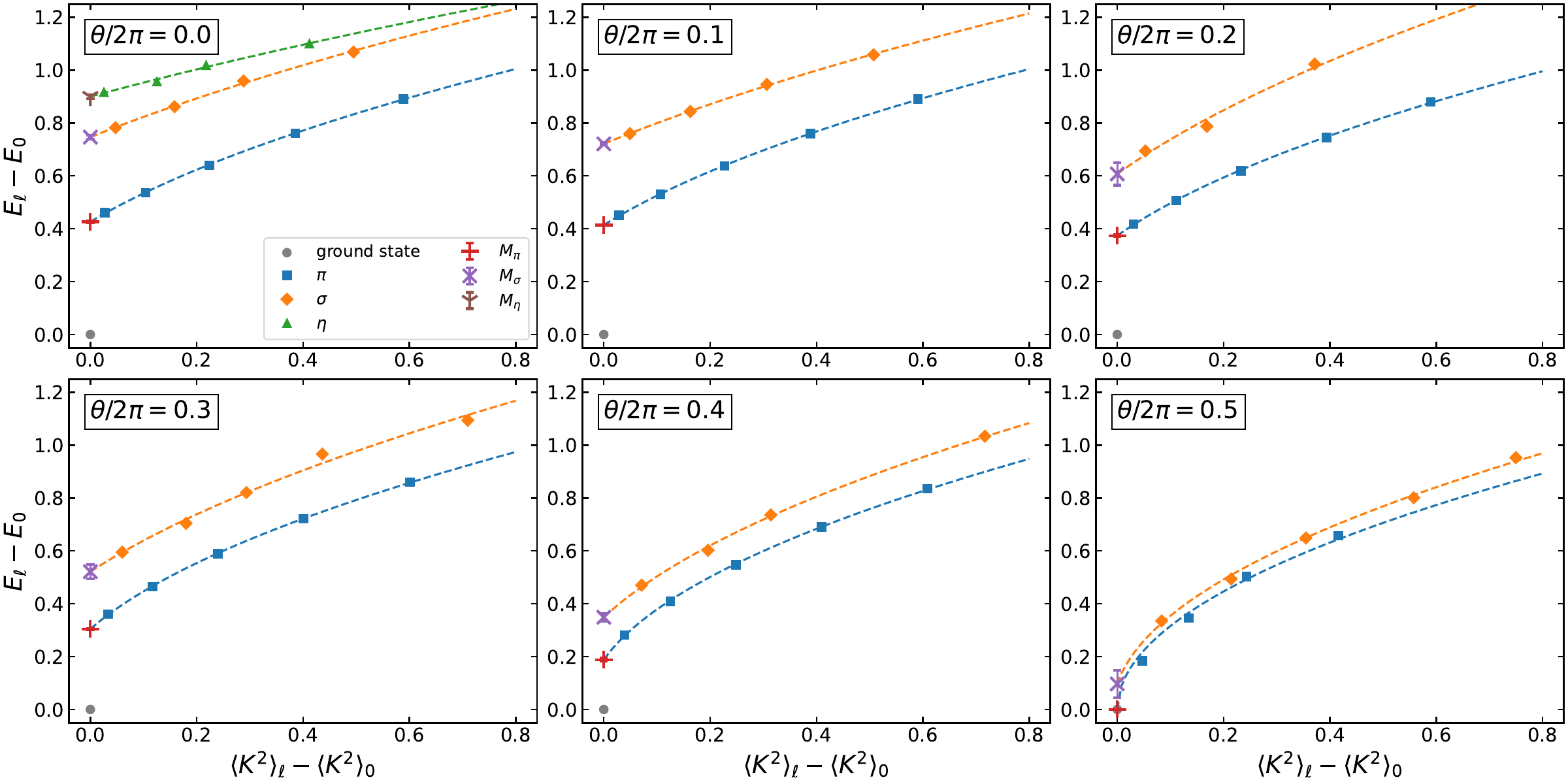}
    \caption{\label{fig_E_vs_K2} 
    The energy gap $\Delta E_{\ell}$ is plotted against the square of total momentum $\Delta K_{\ell}^2$ for each meson.
    The dashed lines depict the fitting results of the dispersion relations. 
    The resulting meson masses are indicated by the markers at the left endpoints with error bars.}
\end{figure}

In the calculation of the dispersion-relation scheme, 
we set the lattice size to $N=100$, the lattice spacing to $a=0.2$, and the weight in eq.~\eqref{eq_H_ell} to $W=10$.
We generate the energy eigenstates using the Hamiltonian~\eqref{eq_H_ell} with and without the singlet projection, 
and then measure the energy gap $\Delta E_{\ell} = E_{\ell}-E_0$ 
and the momentum square $\Delta K_{\ell}^2 = \braket{K^2}_{\ell}-\braket{K^2}_0$.
Note that the translational invariance is broken by the open boundary in our setup 
and thus $\braket{K^2}$ is not a genuine quantum number. 
Nevertheless, it works well empirically as an approximation by subtracting the ground-state contribution $\braket{K^2}_0$.

We identify the series of triplets with the monotonically increasing momentum as the pions, which have the isospin $J=1$.
Similarly, we identify the lowest series with $J=0$ as the sigma mesons.
At $\theta = 0$, we also find the eta meson with $G < 0$.
For $\theta \neq 0$, the eta meson is no longer stable due to the absence of the $G$-parity, 
and thus the corresponding states are replaced by scattering states. 
We plot the energy gap $\Delta E_{\ell}$ against the momentum square $\Delta K_{\ell}^2$ 
for each meson in figure~\ref{fig_E_vs_K2}.
Then we fit these data points by $\Delta E = \sqrt{b^2 \Delta K^2 + M^2}$ with parameters $M$ and $b$.
The result of $M$ is regarded as the meson mass as the extrapolation to $\Delta K^2 \to 0$.

\section{Discussion}
\label{sec_discussion}

We investigated the $\theta$-dependent mass spectrum of the 2-flavor Schwinger model in the lattice Hamiltonian formalism.
The one-point-function and dispersion-relation schemes were developed in our previous work~\cite{Itou:2023img} 
and have now been extended to the case of $\theta \neq 0$~\cite{Itou:2024psm}.
These two schemes can reproduce the mass spectrum even for large $\theta$ without the sign problem, 
which was not possible by the Monte Carlo study with the reweighting technique~\cite{Fukaya:2003ph}.

Our numerical results at the finite fermion mass $m/g = 0.1$ 
agree with the analytic result of the bosonized model that assumes $m/g \ll 1$.
The mechanism of this agreement and the range of applicability remain theoretical questions.
The 2-flavor Schwinger model at $\theta = \pi$ shows nearly conformal behavior, 
where the mass gap is invisibly small for the numerical study~\cite{Coleman:1976uz, Dempsey:2023gib}.
In this case, the DMRG results of the one-point functions are consistent with the CFT calculation, 
indicating that the model at $\theta = \pi$ is well approximated by the $\mathrm{SU}(2)_1$ WZW model.

We obtained promising results thanks to the efficient tensor network algorithm (DMRG) in the $(1+1)$d.
For the extension to higher dimensions, implementing our method on a quantum computer would be an important prospect.
It is also interesting to apply the method to finite-density systems, 
which suffer from the sign problem in the conventional Monte Carlo method.

\acknowledgments
We would like to thank S.~Aoki, M.~Honda, K.~Murakami, S.~Takeda, and A.~Ueda for their useful discussions.
The numerical calculations were carried out on Yukawa-21 at YITP in Kyoto University and the PC clusters at RIKEN iTHEMS.
The work of E.~I. is supported by JST PRESTO Grant Number JPMJPR2113, 
JSPS Grant-in-Aid for Transformative Research Areas (A) JP21H05190, 
JST Grant Number JPMJPF2221,  
JPMJCR24I3,  
and also supported by Program for Promoting Researches on the Supercomputer ``Fugaku'' (Simulation for basic science: from fundamental laws of particles to creation of nuclei) and (Simulation for basic science: approaching the new quantum era), and Joint Institute for Computational Fundamental Science (JICFuS), Grant Number JPMXP1020230411. 
The work of Y.~T. is supported by Japan Society for the Promotion of Science (JSPS) KAKENHI Grant No. 23K22489.
This work is supported by Center for Gravitational Physics and Quantum Information (CGPQI) at Yukawa Institute for Theoretical Physics.

\bibliographystyle{JHEP}
\bibliography{Nf2_Schwinger.bib, QFT.bib}

\end{document}